\documentclass[aps,prb,amsmath,twocolumn]{revtex4}
\usepackage{bm}
\usepackage{graphicx}
\DeclareMathOperator{\Sp}{\mathrm{Sp}}
\begin{document}

\title{Crossed Andreev reflection at spin-active interfaces}
\author{Mikhail S. Kalenkov}
\affiliation{I.E. Tamm Department of Theoretical Physics, P.N.
Lebedev Physics Institute, 119991 Moscow, Russia}
\author{Andrei D. Zaikin}
\affiliation{Forschungszentrum Karlsruhe, Institut f\"ur Nanotechnologie,
76021, Karlsruhe, Germany}
\affiliation{I.E. Tamm Department of Theoretical Physics, P.N.
Lebedev Physics Institute, 119991 Moscow, Russia}

\begin{abstract}
With the aid of the quasiclassical Eilenberger formalism we
develop a theory of non-local electron transport across
three-terminal ballistic normal-superconducting-normal (NSN)
devices with spin-active NS interfaces. The phenomenon of crossed
Andreev reflection (CAR) is known to play the key role in such
transport. We demonstrate that CAR is highly sensitive to electron
spins and yields a rich variety of properties of  non-local
conductance which we describe non-perturbatively at arbitrary
voltages, temperature, spin-dependent interface transmissions and
their polarizations. Our results can be applied to multi-terminal
hybrid structures with normal, ferromagnetic and half-metallic
electrodes and can be directly tested in future experiments.
\end{abstract}

\maketitle

\section{Introduction}

Low energy electron transport in hybrid structures composed of a
normal metal (N) and a superconductor (S) is governed by Andreev
reflection \cite{And} (AR) which causes non-zero subgap
conductance \cite{BTK} of such structures. AR remains essentially
a local effect provided there exists only one NS interface in the
system or, else, if the distance between different NS interfaces
greatly exceeds the superconducting coherence length $\xi$. If,
however, the distance $L$ between two adjacent NS interfaces (i.e.
the superconductor size) is smaller than (or comparable with)
$\xi$, two additional {\it non-local} processes come into play
(see Fig. 1). One such process corresponds to direct electron
transfer between two N-metals through a superconductor. Another
process is the so-called crossed Andreev reflection \cite{BF,GF}
(CAR): An electron penetrating into the superconductor from the
first N-terminal may form a Cooper pair together with another
electron from the second N-terminal. In this case a hole will go
into the second N-metal and AR becomes a non-local effect. Both
these processes  contribute to the non-local conductance of hybrid
multi-terminal structures which has been directly measured in
several recent experiments \cite{Beckmann,Teun,Venkat}.

\begin{figure}
\includegraphics[width=75mm]{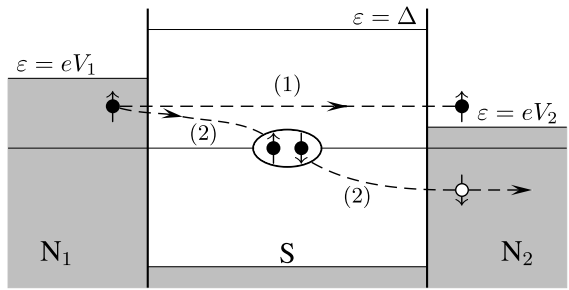}
\caption{Two elementary processes contributing to non-local
conductance of an NSN device: (1) direct electron transfer and (2)
crossed Andreev reflection.}
\end{figure}

Theoretically the non-local conductance of NSN hybrids was
analyzed within the perturbation theory in the transmission of NS
interfaces in Refs. \onlinecite{FFH,Fabio} where it was
demonstrated that in the lowest order in the interface
transmission and at $T=0$ CAR contribution to cross-terminal
conductance is exactly canceled by that from elastic electron
cotunneling (EC), i.e. the non-local conductance vanishes in this
limit. Thus, in order to determine the scale of the effect it is
necessary to include higher order (in the barrier transmission)
terms into consideration. The corresponding analysis was employed
in Refs. \onlinecite{MF,Melin} by means of effective ``dressing''
of both EC and CAR contributions by higher order processes and
more recently by the present authors \cite{KZ06} with the
framework of quasiclassical formalism of Eilenberger equations. We
note that the results \cite{MF} and \cite{KZ06} disagree beyond
perturbation theory since  ``dressing'' procedure \cite{MF} does
not account for all higher order processes. Hence, the approach
\cite{MF} is in general insufficient to correctly describe
non-trivial interplay between normal reflection, tunneling, local
AR and CAR to all orders in the interface transmissions. We will
return to this issue further below.

Another interesting issue is the effect of disorder. It is well
known that disorder enhances interference effects and, hence, can
strongly modify local subgap conductance of NS interfaces in the
low energy limit \cite{VZK,HN,Z}. Non-local conductance of
multi-terminal hybrid NSN structures in the presence of disorder
was recently studied in Refs. \onlinecite{BG,Belzig,Duhot,GZ07}.
Brinkman and Golubov made use of the quasiclassical formalism of
Usadel equations and proceeded perturbatively in the interface
transmissions. Duhot and Melin \cite{Duhot} discussed
the impact of weak-localization-type of effects inside the superconductor on
non-local electron transport in NSN structures.
Morten {\it et al.} \cite{Belzig} employed the
circuit theory (thereby going beyond perturbation theory in
tunneling) and considered a device with normal terminals attached
to a superconductor via an additional {\it normal} island (dot)
\cite{FN}. Very recently a similar structure with a {\it
superconducting} dot was analyzed \cite{GZ07} providing a rather
general theoretical framework to study non-local electron
transport in multi-terminal NSN structures in the presence of
disorder and non-equilibrium
effects.

Yet another interesting subject is an interplay between CAR and
Coulomb interaction. The effect of electron-electron interactions
on AR was investigated in a number of papers \cite{Z,HHK,GZ06}.
Interactions should also affect CAR, e.g., by lifting the exact
cancellation of EC and CAR contributions \cite{LY} already in the
lowest order in tunneling. A complete theory of
non-local transport in realistic NSN systems should include both
disorder and interactions which remains an important task for
future investigations.

An important property of both AR and CAR is that these processes
should be sensitive to magnetic properties of normal electrodes
because these processes essentially depend on spins of scattered
electrons. One possible way to demonstrate spin-resolved CAR is to
use ferromagnets (F) instead of normal electrodes
ferromagnet-superconductor-ferromagnet (FSF) structures
\cite{Beckmann,MF,Yam,Fazio}. First experiments on such FSF
structures \cite{Beckmann} illustrated this point by demonstrating
the dependence of non-local conductance on the polarization of
ferromagnetic terminals. Hence, for better understanding of
non-local effects in multi-terminal hybrid proximity structures it
is desirable to construct a theory of {\it spin-resolved}
non-local transport. In the lowest order in tunneling this
task was accomplished in Ref. \onlinecite{FFH}. For FSF structures
higher orders in the interface transmissions were considered in
Refs. \cite{MF,Melin}.

In this paper we are going to generalize our quasiclassical
approach \cite{KZ06} and construct a theory of non-local electron
transport in ballistic NSN structures with spin-active interfaces
to all orders in their transmissions. Our model allows to
distinguish spin-dependent contributions to the non-local
conductance and to effectively mimic the situation of
ferromagnetic and/or half-metallic electrodes.

The structure of the paper is as follows. In Sec. 2 we introduce
our model and discuss the quasiclassical formalism supplemented by
the boundary conditions for Green-Keldysh functions which account
for electron scattering at spin-active interfaces. In Sec. 3 we
employ this formalism and develop a theory of non-local
spin-resolved electron transport in NSN structures with
spin-active interfaces. Our main conclusions are briefly
summarized in Sec. 4.

\section{The model and formalism}
\begin{figure}
\centerline{\includegraphics[width=65mm]{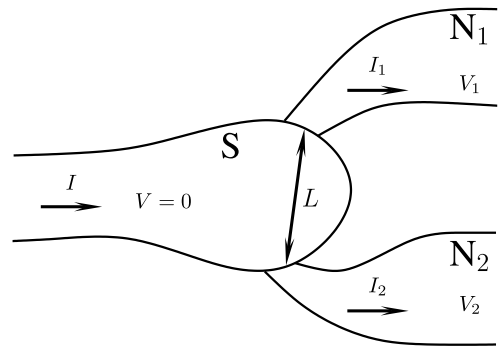}}
\caption{Schematics of our NSN device.} \label{nsn}
\end{figure}

Let us consider three-terminal NSN structure depicted in Fig.
\ref{nsn}. We will assume that all three metallic electrodes are
non-magnetic and ballistic, i.e. the electron elastic mean free
path in each metal is larger than any other relevant size scale.
In order to resolve spin-dependent effects we will assume that
both NS interfaces are spin-active, i.e. we will distinguish
``spin-up'' and ``spin-down'' transmissions of the first
($D_{1\uparrow}$ and $D_{1\downarrow}$) and the second
($D_{2\uparrow}$ and $D_{2\downarrow}$) SN interface. All these
four transmissions may take any value from zero to one. We also
introduce the angle $\varphi$ between polarizations of two
interfaces which can take any value between 0 and $2\pi$.

In what follows effective cross-sections of the two interfaces
will be denoted respectively as $\mathcal{A}_1$ and
$\mathcal{A}_2$. The distance between these interfaces $L$ as well
as other geometric parameters are assumed to be much larger than
$\sqrt{\mathcal{A}_{1,2}}$, i.e. effectively both contacts are
metallic constrictions. In this case the voltage drops only across
SN interfaces and not inside large metallic electrodes. Hence,
nonequilibrium (e.g. charge imbalance) effects related to the
electric field penetration into the S-electrode can be neglected.
In our analysis we will also disregard Coulomb effects
\cite{Z,HHK,GZ06}.

For convenience, we will set the electric potential of the
S-electrode equal to zero, $V=0$. In the presence of bias voltages
$V_1$ and $V_2$ applied to two normal electrodes (see Fig.
\ref{nsn}) the currents $I_1$ and $I_2$ will flow through SN$_1$
and SN$_2$ interfaces. These currents can be evaluated with the
aid of the quasiclassical formalism of nonequilibrium
Green-Eilenberger-Keldysh functions \cite{BWBSZ} $\hat g^{R,A,K}$
which we briefly specify below.

\subsection{Quasiclassical equations}

In the ballistic limit the corresponding Eilenberger equations
take the form
\begin{gather}
\begin{split}
\left[
\varepsilon \hat\tau_3+
eV(\bm{r},t)-
\hat\Delta(\bm{r},t),
\hat g^{R,A,K} (\bm{p}_F, \varepsilon, \bm{r},t)
\right]
+\\+
i\bm{v}_F \nabla \hat g^{R,A,K} (\bm{p}_F, \varepsilon, \bm{r},t) =0,
\end{split}
\label{Eil}
\end{gather}
where $[\hat a, \hat b]= \hat a\hat b - \hat b \hat
a$, $\varepsilon$ is the quasiparticle energy, $\bm{p}_F=m\bm{v}_F$ is the
electron Fermi momentum vector and $\hat\tau_3$ is the Pauli matrix in Nambu
space.
The functions  $\hat g^{R,A,K}$ also obey the normalization conditions
$(\hat g^R)^2=(\hat g^A)^2=1$ and $\hat g^R \hat g^K + \hat g^K \hat g^A =0$.
Here and below the product of matrices is defined as time convolution.

Green functions $\hat g^{R,A,K}$ and $\hat\Delta$ are $4\times4$ matrices in
Nambu and spin spaces. In Nambu space they can be parameterized as
\begin{equation}
        \hat g^{R,A,K} =
        \begin{pmatrix}
                g^{R,A,K} & f^{R,A,K} \\
                \tilde f^{R,A,K} & \tilde g^{R,A,K} \\
        \end{pmatrix}, \quad
        \hat\Delta=
        \begin{pmatrix}
                0 & \Delta i\sigma_2 \\
                \Delta^* i\sigma_2& 0 \\
        \end{pmatrix},
\end{equation}
where $g^{R,A,K}$, $f^{R,A,K}$, $\tilde f^{R,A,K}$, $\tilde g^{R,A,K}$ are
$2\times2$ matrices in the spin space, $\Delta$ is the BCS order parameter and
$\sigma_i$ are Pauli matrices. For simplicity we will only consider the case of
spin-singlet isotropic pairing in the superconducting electrode.
The current density is
related to the Keldysh function $\hat g^K$ according to the standard relation
\begin{equation}
\bm{j}(\bm{r}, t)= -\dfrac{e N_0}{8} \int d \varepsilon
\left< \bm{v}_F \mathrm{Sp} [\hat \tau_3 \hat g^K(\bm{p}_F,
\varepsilon, \bm{r},t)] \right>,
\label{current}
\end{equation}
where $N_0=mp_F/2\pi^2$ is the density of state at the Fermi level and
angular brackets $\left< ... \right>$ denote averaging over the Fermi momentum.

\subsection{Riccati parameterization}

The above matrix Green-Keldysh functions can be conveniently
parameterized by four Riccati amplitudes $\gamma^{R,A}$, $\tilde
\gamma^{R,A}$ and two ``distribution functions'' $x^K$, $\tilde
x^K$ (here and below we chose to follow the notations
\cite{Eschrig00}):
\begin{equation}
\hat g^K=
2
\hat N^R
\begin{pmatrix}
x^K - \gamma^R  \tilde x^K  \tilde \gamma^A &
-\gamma^R  \tilde x^K + x^K  \gamma^A \\
-\tilde \gamma^R  x^K + \tilde x^K  \tilde \gamma^A &
\tilde x^K - \tilde \gamma^R  x^K  \gamma^A \\
\end{pmatrix}
\hat N^A,
\label{gkparam}
\end{equation}
where functions $\gamma^{R,A}$ and $\tilde \gamma^{R,A}$ are Riccati amplitudes
\begin{equation}
\hat g^{R,A}=\pm
    \hat N^{R,A}
    \begin{pmatrix}
    1+\gamma^{R,A} \tilde \gamma^{R,A} & 2\gamma^{R,A} \\
    -2 \tilde \gamma^{R,A} & -1- \tilde \gamma^{R,A}  \gamma^{R,A} \\
    \end{pmatrix}
    \label{graparam}
\end{equation}
and $\hat N^{R,A}$ are the following matrices
\begin{equation}
\hat N^{R,A}=
    \begin{pmatrix}
    (1-\gamma^{R,A} \tilde \gamma^{R,A})^{-1} & 0 \\
    0 & (1-\tilde \gamma^{R,A}  \gamma^{R,A} )^{-1} \\
    \end{pmatrix}.
    \label{nrparam}
\end{equation}
With the aid of the above parameterization one can identically transform
the quasiclassical equations \eqref{Eil} into the following set of
effectively decoupled equations for
Riccati amplitudes and distribution functions \cite{Eschrig00}
\begin{gather}
\begin{split}
i\bm{v}_F \nabla \gamma^{R,A}+ [\varepsilon+eV(\bm{r},t)]\gamma^{R,A}+
\gamma^{R,A}[\varepsilon-eV(\bm{r},t)]
\\
=\gamma^{R,A}\Delta^* i\sigma_2\gamma^{R,A}-\Delta i\sigma_2,
\end{split}
\label{eqgamma}
\\
\begin{split}
i\bm{v}_F \nabla \tilde\gamma^{R,A}- [\varepsilon-eV(\bm{r},t)]\tilde\gamma^{R,A}-
\tilde\gamma^{R,A}[\varepsilon+eV(\bm{r},t)]
\\
=\tilde\gamma^{R,A}\Delta i\sigma_2\tilde\gamma^{R,A}-\Delta^* i\sigma_2,
\end{split}
\label{eqtildegamma}
\\
\begin{split}
i\bm{v}_F \nabla
x^K+[\varepsilon+eV(\bm{r},t)]x^K-x^K[\varepsilon+eV(\bm{r},t)]
\\
-\gamma^{R}\Delta^* i\sigma_2 x^K-
x^K \Delta i\sigma_2 \tilde\gamma^{A}=0,
\end{split}
\label{eqx}
\\
\begin{split}
i\bm{v}_F \nabla \tilde x^K-
[\varepsilon-eV(\bm{r},t)]\tilde x^K+\tilde x^K[\varepsilon-eV(\bm{r},t)]
\\
-\tilde \gamma^{R}\Delta i\sigma_2 \tilde x^K-
\tilde x^K \Delta^* i\sigma_2 \gamma^{A}=0.
\end{split}
\label{eqtildex}
\end{gather}

Depending on the particular trajectory it is also convenient to
introduce a ``replica'' of both Riccati amplitudes and
distribution functions which -- again following the notations
\cite{Eschrig00,Zhao04} -- will be denoted by capital letters
$\Gamma$ and $X$. These ``capital'' Riccati amplitudes and
distribution functions obey the same equations
\eqref{eqgamma}-\eqref{eqtildex} with the replacement $\gamma
\rightarrow \Gamma$ and $x \rightarrow X$. The distinction between
different Riccati amplitudes and distribution functions will be
made explicit below.

\subsection{Boundary conditions}

\begin{figure}
\centerline{\includegraphics[width=75mm]{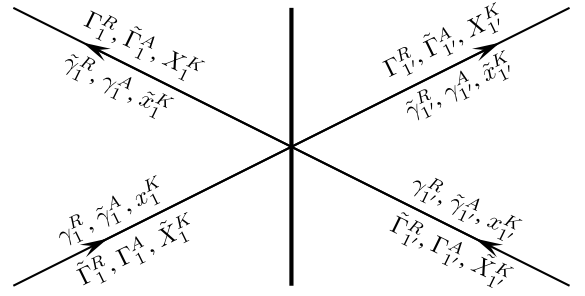}} \caption{Riccati
amplitudes for incoming and outgoing trajectories from the both
sides of the interface. } \label{sis}
\end{figure}

Quasiclassical equations should be supplemented by appropriate
boundary conditions at metallic interfaces. In the case of
specularly reflecting spin-degenerate interfaces these conditions
were derived by Zaitsev \cite{Zaitsev} and later generalized to
spin-active interfaces in Ref. \onlinecite{Millis88}.

Before specifying these conditions it is important to emphasize
that the applicability of the Eilenberger quasiclassical formalism
with appropriate boundary conditions to hybrid structures with two
or more barriers is, in general, a non-trivial issue
\cite{GZ02,OS}. Electrons scattered at different barriers
interfere and form bound states (resonances) which cannot be
correctly described within the quasiclassical formalism employing
Zaitsev boundary conditions or their direct generalization. Here
we avoid this problem by choosing the appropriate geometry of our
NSN device, see Fig. 2. In our system any relevant trajectory
reaches each NS interface only once whereas the probability of
multiple reflections at both interfaces is small in the parameter
$\mathcal{A}_1 \mathcal{A}_2/L^4 \ll 1$. Hence, resonances formed
by multiply reflected electron waves can be neglected, and our
formalism remains adequate for the problem in question.

It will be convenient for us to formulate the boundary conditions
directly in terms of Riccati amplitudes and the distribution
functions. Let us consider the first NS interface and explicitly
specify the relations between Riccati amplitudes and distribution
functions for incoming and outgoing trajectories, see
Fig.~\ref{sis}. The boundary conditions for $\Gamma_1^R$,
$\tilde\Gamma_1^A$ and $X_1^K$ can be written in the form
\cite{Zhao04}
\begin{gather}
\Gamma_1^R= r_{1l}^R\gamma_1^R \underline{S}_{11}^+ +
t_{1l}^R\gamma_{1'}^R {\underline{S}}_{11'}^+,
\\
\tilde\Gamma_1^A= \underline{S}_{11} \tilde\gamma_1^A \tilde r_{1r}^A  +
{\underline{S}}_{11'} \tilde\gamma_{1'}^A \tilde t_{1r}^A ,
\\
X_1^K=r_{1l}^R x_1^K \tilde r_{1r}^A +
t_{1l}^R x_{1'}^K \tilde t_{1r}^A -
a_{1l}^R \tilde x_{1'}^K \tilde a_{1r}^A.
\end{gather}
Here we defined the transmission ($t$), reflection  ($r$), and
branch-conversion ($a$) amplitudes as:
\begin{gather}
r_{1l}^R=+[(\beta_{1'1}^R)^{-1}S_{11}^+ - (\beta_{1'1'}^R)^{-1}S_{11'}^+]^{-1}
(\beta_{1'1}^R)^{-1},
\\
t_{1l}^R=-[(\beta_{1'1}^R)^{-1}S_{11}^+ - (\beta_{1'1'}^R)^{-1}S_{11'}^+]^{-1}
(\beta_{1'1'}^R)^{-1},
\\
\tilde r_{1r}^A=+(\beta_{1'1}^A)^{-1}
[S_{11}(\beta_{1'1}^A)^{-1} - S_{11'}(\beta_{1'1'}^A)^{-1}]^{-1},
\\
\tilde t_{1r}^A=-(\beta_{1'1'}^A)^{-1}
[S_{11}(\beta_{1'1}^A)^{-1} - S_{11'}(\beta_{1'1'}^A)^{-1}]^{-1},
\\
a_{1l}^R=(\Gamma_1^R \underline{S}_{11} - S_{11}\gamma_1^R)(\tilde
\beta_{11'}^R)^{-1},
\\
\tilde a_{1r}^A=(\tilde \beta_{11'}^A)^{-1}
(\underline{S}_{11}^+ \tilde \Gamma_1^A  - \tilde \gamma_1^A S_{11}^+),
\end{gather}
where
\begin{gather}
\beta_{ij}^R=S_{ij}^+ - \gamma_j^R \underline{S}_{ij}^+ \tilde\gamma_i^R,\
\tilde\beta_{ij}^R=\underline{S}_{ji} - \tilde \gamma_j^R S_{ji}\gamma_i^R,
\\
\beta_{ij}^A=S_{ij} - \gamma_i^A \underline{S}_{ij} \tilde\gamma_j^A,\
\tilde\beta_{ij}^A=\underline{S}_{ji}^+ - \tilde \gamma_i^A S_{ji}^+\gamma_j^A.
\end{gather}
Similarly, the boundary conditions for $\tilde\Gamma_1^R$,
$\Gamma_1^A$, and $\tilde X_1^K$ take the form:
\begin{gather}
\tilde\Gamma_1^R= \tilde r_{1l}^R\tilde\gamma_1^R S_{11} +
\tilde t_{1l}^R\tilde\gamma_{1'}^R S_{1'1},
\\
\Gamma_1^A= S_{11}^+ \gamma_1^A r_{1r}^A  +
S_{1'1}^+ \gamma_{1'}^A t_{1r}^A ,
\\
\tilde X_1^K=\tilde r_{1l}^R \tilde x_1^K r_{1r}^A +
\tilde t_{1l}^R \tilde x_{1'}^K t_{1r}^A -
\tilde a_{1l}^R x_{1'}^K a_{1r}^A,
\end{gather}
where
\begin{gather}
\tilde r_{1l}^R=+[(\tilde \beta_{1'1}^R)^{-1}\underline{S}_{11} -
(\tilde \beta_{1'1'}^R)^{-1}\underline{S}_{1'1}]^{-1}
(\tilde\beta_{1'1}^R)^{-1},
\\
t_{1l}^R=-[(\tilde \beta_{1'1}^R)^{-1}\underline{S}_{11} -
(\tilde \beta_{1'1'}^R)^{-1}\underline{S}_{1'1}]^{-1}
(\tilde\beta_{1'1'}^R)^{-1},
\\
r_{1r}^A=+(\tilde\beta_{1'1}^A)^{-1}
[\underline{S}_{11}^+(\tilde\beta_{1'1}^A)^{-1} -
\underline{S}_{1'1}^+(\tilde\beta_{1'1'}^A)^{-1}]^{-1},
\\
\tilde t_{1r}^A=-(\tilde\beta_{1'1'}^A)^{-1}
[\underline{S}_{11}^+(\tilde\beta_{1'1}^A)^{-1} -
\underline{S}_{1'1}^+(\tilde\beta_{1'1'}^A)^{-1}]^{-1},
\\
\tilde a_{1l}^R=(\tilde \Gamma_1^R S_{11}^+ - \underline{S}_{11}^+ \tilde \gamma_1^R)( \beta_{11'}^R)^{-1},
\\
a_{1r}^A=(\beta_{11'}^A)^{-1} (S_{11} \Gamma_1^A  - \gamma_1^A \underline{S}_{11}).
\end{gather}
Boundary conditions for $\Gamma_{1'}^{R,A}$, $\tilde
\Gamma_{1'}^{R,A}$, $X^K_{1'}$ and $\tilde X^K_{1'}$ can be
obtained from the above equations simply by replacing $1
\leftrightarrow 1'$.

The matrices $S_{11}$, $S_{11'}$, $S_{1'1}$, and $S_{1'1'}$
constitute the components of the $\mathcal{S}$-matrix describing
electron scattering at the first interface:
\begin{equation}
\mathcal{S}=
\begin{pmatrix}
S_{11} & S_{11'}\\
S_{1'1} & S_{1'1'}\\
\end{pmatrix}, \quad
\mathcal{S}\mathcal{S}^+=1
\end{equation}

\begin{figure*}
\centerline{
\includegraphics{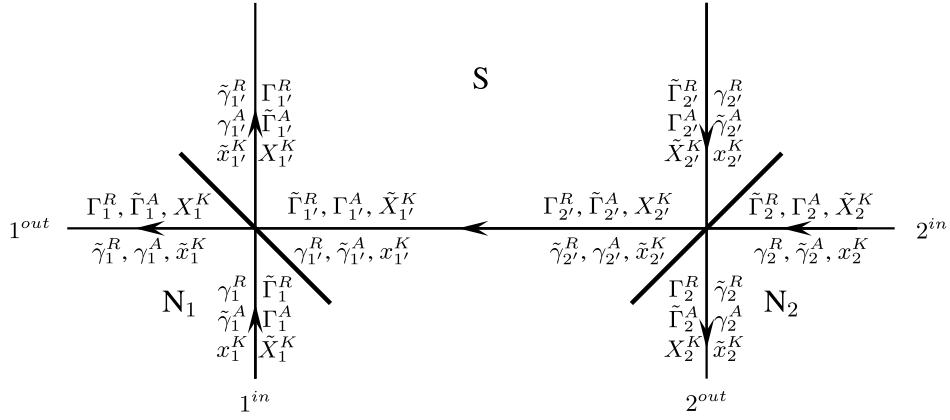}
} \caption{Riccati amplitudes for incoming and outgoing
trajectories for an NSN structure with two barriers. The arrows
define quasiparticle momentum directions. We also indicate
relevant Riccati amplitudes and distribution functions
parameterizing the Green-Keldysh function for the corresponding
trajectory.} \label{traject_b}
\end{figure*}

In our three terminal geometry nonlocal conductance arises only
from trajectories that cross both interfaces, as illustrated in
Fig.~\ref{traject_b}. Accordingly, the above boundary conditions
should be employed at both NS interfaces.

Finally, one needs to specify the asymptotic boundary conditions
far from NS interfaces. Deep in metallic electrodes we have
\begin{gather}
\gamma_1^R=\tilde \gamma_1^R=\gamma_1^A=\tilde \gamma_1^A=0,
\\
x_1^K=h_0(\varepsilon+eV_1),\quad
\tilde x_1^K=-h_0(\varepsilon-eV_1),
\\
\gamma_2^R=\tilde \gamma_2^R=\gamma_2^A=\tilde \gamma_2^A=0,
\\
x_2^K=h_0(\varepsilon+eV_2),\quad
\tilde x_2^K=-h_0(\varepsilon-eV_2),
\end{gather}
where $h_0(\varepsilon)=\tanh(\varepsilon/2T)$ - equilibrium distribution function.
In the bulk of superconducting electrode we have
\begin{gather}
\tilde \gamma_{1'}^R=-a(\varepsilon)i\sigma_2, \quad
\gamma_{1'}^A=a^*(\varepsilon)i\sigma_2,
\\
\tilde x_{1'}^K=-[1-|a(\varepsilon)|^2]h_0(\varepsilon),
\\
\gamma_{2'}^R=a(\varepsilon)i\sigma_2, \quad
\tilde\gamma_{2'}^A=-a^*(\varepsilon)i\sigma_2,
\\
x_{2'}^K=[1-|a(\varepsilon)|^2]h_0(\varepsilon),
\end{gather}
where we denoted $a( \varepsilon ) = - (\varepsilon - \sqrt{
\varepsilon^2 - \Delta^2})/\Delta$.

\subsection{Green functions}

With the aid of the above equations and boundary conditions it is
straightforward to evaluate the quasiclassical Green-Keldysh
functions for our three-terminal device along any trajectory of
interest. For instance, from the boundary conditions at the second
interface we find
\begin{equation}
\Gamma_{2'}^R=ia(\varepsilon)A_2\sigma_2,
\label{GGG}
\end{equation}
where $A_2=S_{2'2'}\sigma_2\underline{S}_{2'2'}^+ \sigma_2$.
Integrating Eq. \eqref{eqgamma} along the trajectory connecting
both interfaces and using Eq.~(\ref{GGG}) as the initial condition
we immediately evaluate the Riccati amplitude at the first
interface:
\begin{gather}
\gamma_{1'}^R=
i\dfrac{aA_2 + (aA_2\varepsilon + \Delta)Q
}{1-(aA_2\Delta + \varepsilon)Q}\sigma_2,
\\
Q=\dfrac{\tanh\left[i\Omega L/v_F\right]}{
\Omega}, \quad
\Omega = \sqrt{\varepsilon^2-\Delta^2}.
\end{gather}
Employing the boundary conditions again we obtain
\begin{gather}
\Gamma_1^R=iS_{11'}K_{21}^{-1}
\left[aA_2+(aA_2\varepsilon + \Delta)Q\right]
\sigma_2\underline{S}_{11'}^+,
\\
\tilde\Gamma_1^R=-i a \underline{S}_{1'1}^+ \sigma_2 S_{1'1'}
K_{21}^{-1}\left[1-(aA_2\Delta+\varepsilon)Q\right]
S_{1'1'}^{-1} S_{1'1},
\end{gather}
where
\begin{gather}
K_{ij}= (1-a^2 A_i A_j) - \left[ \varepsilon (1+a^2 A_i A_j) + \Delta a (A_i+A_j) \right]Q,
\\
A_1=\sigma_2 \underline{S}_{1'1'}^+\sigma_2 S_{1'1'}.
\end{gather}
We also note that the relation $(\Gamma^{R,A})^+ = \tilde \Gamma^{A,R}$ and $(\gamma^{R,A})^+ = \tilde \gamma^{A,R}$
makes it unnecessary (while redundant) to separately
calculate the advanced Riccati amplitudes.

Let us now evaluate the distribution functions at both interfaces.
With the aid of the boundary conditions at the second interface we
obtain
\begin{multline}
X_{2'}^K=
S_{2'2'}S_{2'2'}^+\left(1-|a|^2\right)h_0(\varepsilon) +
S_{2'2}S_{2'2}^+ x_2^K
-\\-
|a|^2 S_{2'2'} \sigma_2 \underline{S}_{22'}^+\underline{S}_{22'} \sigma_2
S_{2'2'}^+ \tilde x_2^K.
\end{multline}
Integrating Eq. \eqref{eqx} along the trajectory connecting both
interfaces with initial condition for $X_{2'}^K$, we arrive at the
expression for $x_{1'}^K$
\begin{multline}
x_{1'}^K=
\left[1-(a A_2\Delta + \varepsilon)Q\right]^{-1} X_{2'}^K
\times\\\times
(1-\tanh^2iL\Omega/v_F)
\left[1-(aA_2\Delta + \varepsilon)Q\right]^{+^{-1}}.
\end{multline}
Then we can find distribution functions at the first interface. On
the normal metal side of the interface we find
\begin{equation}
X_1^K=
r_{1l}^R x_1^K r_{1l}^{R^+} +
t_{1l}^R x_{1'}^K t_{1l}^{R^+} +
a_{1l}^R  a_{1l}^{R^+} \left(1-|a|^2\right)h_0(\varepsilon)
\end{equation}
where
\begin{gather}
\begin{split}
r_{1l}^R=S_{11'}K_{21}^{-1}
\Bigl[ (1-(aA_2\Delta + \varepsilon)Q)S_{1'1'}^+S_{1'1}^{+^{-1}}
-\\-
a(aA_2+(aA_2\varepsilon + \Delta)Q) \sigma_2 \underline{S}_{1'1'}^+ \sigma_2 S_{1'1}^{+^{-1}} \Bigr],
\end{split}
\\
t_{1l}^R=S_{11'}K_{21}^{-1} (1-(aA_2\Delta + \varepsilon)Q),
\\
a_{1l}^R=i S_{11'}K_{21}^{-1}(aA_2+(aA_2\varepsilon + \Delta)Q)\sigma_2 \underline{S}_{1'1'}^+.
\end{gather}
The corresponding expression for $\tilde X_1^K$ is obtained
analogously. We get
\begin{equation}
\tilde X_1^K= \tilde r_{1l}^R \tilde x_1^K \tilde r_{1l}^{R^+} -
\tilde t_{1l}^R \tilde t_{1l}^{R^+} \left(1-|a|^2\right)h_0(\varepsilon) -
\tilde a_{1l}^R x_{1'}^K \tilde a_{1l}^{R^+}.
\end{equation}
where
\begin{gather}
\begin{split}
\tilde r_{1l}^R=-
\Bigl[ \underline{S}_{1'1}^{-1}\underline{S}_{1'1'} \sigma_2 (1-(aA_2\Delta + \varepsilon)Q)
-\\-
\underline{S}_{1'1}^{-1} \sigma_2 S_{1'1'} a(aA_2+(aA_2\varepsilon + \Delta)Q)  \Bigr]
K_{12}^{-1} \sigma_2 \underline{S}_{11'}^+,
\end{split}
\\
\tilde t_{1l}^R=\underline{S}_{1'1}^+ \underline{S}_{1'1'}^{+^{-1}} \sigma_2
(1-(aA_2\Delta + \varepsilon)Q) K_{12}^{-1}
\sigma_2 \underline{S}_{1'1'}^+,
\\
\tilde a_{1l}^R= i a \underline{S}_{1'1}^+ \sigma_2 S_{1'1'}
 K_{21}^{-1} (1-(aA_2\Delta + \varepsilon)Q).
\end{gather}
Combining the above results for the Riccati amplitudes and the
distribution functions we can easily evaluate the Keldysh Green
function at the first interface. For instance, for the trajectory
$1^{out}$ (see Fig.~\ref{traject_b}) we obtain
\begin{equation}
g^K_{1^{out}}=2(X_1^K - \Gamma_1^R \tilde x_1^K \Gamma_1^{R^+}), \quad
\tilde g^K_{1^{out}}=2\tilde x_1^K.
\end{equation}
The Keldysh Green function for the trajectory $1^{in}$ is
evaluated analogously, and we get
\begin{equation}
g^K_{1^{in}}=2 x_1^K, \quad
\tilde g^K_{1^{in}}=2(\tilde X_1^K-\tilde\Gamma_1^R x_1^K \tilde\Gamma_1^{R^+}).
\end{equation}

\section{Nonlocal conductance}
\subsection{General results}

Now we are ready to evaluate the current $I_1$ across the first
interface. This current takes the form:
\begin{equation}
I_1=I_1^{BTK}(V_1)-\dfrac{G_0}{8e} \int d \varepsilon \Sp
(\hat\tau_3 \hat g^K_{1^{out}} - \hat\tau_3 \hat g^K_{1^{in}}),
\label{current_g}
\end{equation}
where
\begin{equation}
G_0=\frac{8\gamma_1 \gamma_2
\mathcal{N}_1\mathcal{N}_2}{R_qp_F^2L^2}
\end{equation}
is the normal state nonlocal conductance of our device at fully transparent
interfaces, $p_F\gamma _{1(2)}$ is normal to the first (second)
interface component of the Fermi momentum for electrons
propagating straight between the interfaces,
$\mathcal{N}_{1,2}=p_F^2\mathcal{A}_{1,2}/4\pi$ define the number
of conducting channels of the corresponding interface,
$R_q=2\pi/e^2$ is the quantum resistance unit.

Here $I_1^{BTK}(V_1)$ stands for the contribution to the current
through the first interface coming from trajectories that never
cross the second interface. This is just the standard BTK
contribution \cite{BTK,Zhao04}. The non-trivial contribution is
represented by the last term in Eq. (\ref{current_g}) which
accounts for the presence of the second NS interface. We observe
that this non-local contribution to the current is small as
$\propto 1/p_F^2L^2$ (rather than  $\propto 1/p_F^3L^3$ as suggested
in Ref. \onlinecite{Yam}). This term will be analyzed in details
below.

The functions $\hat g^K_{1^{in}}$ and $\hat g^K_{1^{out}}$ are the
Keldysh Green functions evaluated on the trajectories $1^{in}$ and
$1^{out}$ respectively. Using the above expression for the Riccati
amplitudes and the distribution functions we find
\begin{multline}
\Sp (\hat\tau_3 \hat g^K_{1^{out}} - \hat\tau_3 \hat g^K_{1^{in}})
=\\= 2 \Sp[r_{1l}^R r_{1l}^{R^+} - \tilde \Gamma_1^R  \tilde
\Gamma_1^{R^+} -1] (h_0(\varepsilon+eV_1)-h_0(\varepsilon)) -\\- 2
\Sp[\tilde r_{1l}^R \tilde r_{1l}^{R^+}- \Gamma_1^R \Gamma_1^{R^+}
-1] (h_0(\varepsilon-eV_1)-h_0(\varepsilon)) +\\+ 2
(1-\tanh^2iL\Omega/v_F) \times\\\times \Sp[K_{21}^{-1}
\{S_{2'2}S_{2'2}^+ (h_0(\varepsilon+eV_2)-h_0(\varepsilon)) +\\+
|a|^2 S_{2'2'} \sigma_2 \underline{S}_{22'}^+\underline{S}_{22'}
\sigma_2 S_{2'2'}^+ (h_0(\varepsilon-eV_2)-h_0(\varepsilon))\}
K_{21}^{+^{-1}} \times\\\times (S_{11'}^+ S_{11'} - |a^2|
S_{1'1'}^{+} \sigma_2 \underline{S}_{1'1} \underline{S}_{1'1}^+
\sigma_2 S_{1'1'}) ], \label{sp_g}
\end{multline}
where we explicitly used the fact that in equilibrium $\Sp
(\hat\tau_3 \hat g^K_{1^{out}} - \hat\tau_3 \hat
g^K_{1^{in}})\equiv 0$. Substituting \eqref{sp_g} into
\eqref{current_g}, we finally obtain
\begin{equation}
I_1= I_1^{BTK}(V_1)+I_{11}(V_1) + I_{12}(V_2).
\label{final}
\end{equation}
The correction to the local BTK current (arising from trajectories
crossing also the second NS interface) has the following form
\begin{multline}
I_{11}(V_1)=
-\dfrac{G_0}{4e} \int d \varepsilon
\bigl\{
\\
\Sp[r_{1l}^R r_{1l}^{R^+} - \tilde \Gamma_1^R  \tilde
\Gamma_1^{R^+} -1] (h_0(\varepsilon+eV_1)-h_0(\varepsilon)) -\\-
\Sp[\tilde r_{1l}^R \tilde r_{1l}^{R^+}- \Gamma_1^R \Gamma_1^{R^+}
-1] (h_0(\varepsilon-eV_1)-h_0(\varepsilon)) \bigr\}, \label{I11}
\end{multline}
while for the cross-current we obtain
\begin{multline}
I_{12}(V_2)=
-\dfrac{G_0}{4e} \int d \varepsilon
(1-\tanh^2iL\Omega/v_F)
\times\\\times
\Sp[K_{21}^{-1}
\{S_{2'2}S_{2'2}^+ (h_0(\varepsilon+eV_2)-h_0(\varepsilon))
+\\+
|a|^2 S_{2'2'} \sigma_2 \underline{S}_{22'}^+\underline{S}_{22'} \sigma_2 S_{2'2'}^+
(h_0(\varepsilon-eV_2)-h_0(\varepsilon))\}
K_{21}^{+^{-1}}
\times\\\times
(S_{11'}^+ S_{11'} -
|a^2| S_{1'1'}^{+} \sigma_2 \underline{S}_{1'1} \underline{S}_{1'1}^+ \sigma_2 S_{1'1'})
].
\label{I12}
\end{multline}
\begin{figure}
\centerline{
\includegraphics[width=85mm]{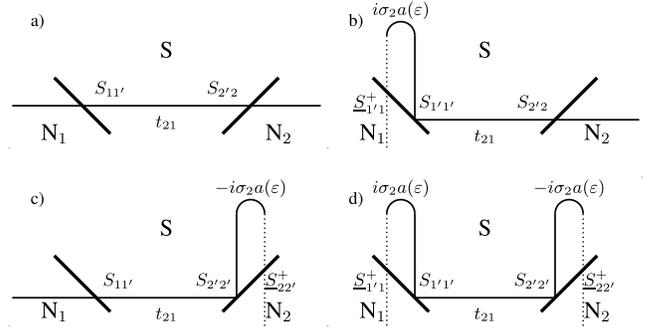}
}
\caption{Diagrams representing four different contributions to the
cross-current $I_{12}$ \eqref{I12}. Solid (dotted) lines correspond to
propagating electron-like (hole-like) excitations and $t_{21}=K_{21}^{-1} / \cosh(iL\Omega/v_F)$.}
\label{traject}
\end{figure}
Eqs. \eqref{final}-\eqref{I12} fully determine the current across
the first interface at arbitrary voltages, temperature and
spin-dependent interface transmissions.

In right hand side of Eq. \eqref{I12} we can distinguish four
contributions with different products of $S$-matrices. Each of
these terms corresponds to a certain sequence of elementary
events, such as transmission, reflection, Andreev reflection and
propagation between interfaces. Diagrammatic representation of
these four terms is offered in Fig. \ref{traject}. The amplitude
of each of the processes is given by the product of the amplitudes
of the corresponding elementary events. For instance, the
amplitude of the process in Fig. 5c  is $f=-i S_{11'}t_{21}
S_{2'2'} a \sigma_2 \underline{S}_{22'}^+$. In  Eq.\eqref{I12}
this process is identified by the term  $\Sp(f f^+)$ with the hole
distribution function as a prefactor. It is straightforward to
observe that the processes of Fig. 5a, 5b and 5d correspond to the
other three terms in \eqref{I12}. We also note that the processes
of Fig. 5a and 5d describe direct electron (hole) transport, while
the processes of Fig. 5b and 5c correspond to the contribution of
CAR.

Assuming that both interfaces possess inversion symmetry as well
as reflection symmetry in the plane normal to the corresponding
interface we can choose $\mathcal{S}$-matrices in the following
form
\begin{multline}
S_{11}=S_{1'1'}=\underline{S}^T_{11}=\underline{S}^T_{1'1'}
=\\=
U(\varphi)
\begin{pmatrix}
\sqrt{R_{1\uparrow}}e^{i\theta_1/2} & 0 \\
0 & \sqrt{R_{1\downarrow}}e^{-i\theta_1/2} \\
\end{pmatrix}
U^+(\varphi),
\label{S11}
\end{multline}
\begin{multline}
S_{11'}=S_{1'1}=\underline{S}^T_{11'}=\underline{S}^T_{1'1} =\\=
U(\varphi) i
\begin{pmatrix}
\sqrt{D_{1\uparrow}}e^{i\theta_1/2} & 0 \\
0 & \sqrt{D_{1\downarrow}}e^{-i\theta_1/2} \\
\end{pmatrix}
U^+(\varphi),
\end{multline}
and
\begin{multline}
S_{22}=S_{2'2'}=\underline{S}_{22}=\underline{S}_{2'2'}
=\\=
\begin{pmatrix}
\sqrt{R_{2\uparrow}}e^{i\theta_2/2} & 0 \\
0 & \sqrt{R_{2\downarrow}}e^{-i\theta_2/2} \\
\end{pmatrix},
\label{S22}
\end{multline}
\begin{multline}
S_{22'}=S_{2'2}=\underline{S}_{22'}=\underline{S}_{2'2}
=\\=
i
\begin{pmatrix}
\sqrt{D_{2\uparrow}}e^{i\theta_2/2} & 0 \\
0 & \sqrt{D_{2\downarrow}}e^{-i\theta_2/2} \\
\end{pmatrix}.
\label{S22'}
\end{multline}
Here $R_{1(2)\uparrow (\downarrow )}=1-D_{1(2)\uparrow (\downarrow
)}$ are the spin dependent reflection coefficients of both NS
interfaces, $\theta_{1,2}$ are spin-mixing angles and $U(\varphi)$
is the rotation matrix in the spin
space which depends on the angle $\varphi$ between polarizations
of the two interfaces,
\begin{equation}
U(\varphi)=\exp(-i\varphi\sigma_1/2)=
\begin{pmatrix}
\cos( \varphi/2) & -i \sin(\varphi/2) \\
-i \sin(\varphi/2) & \cos( \varphi/2) \\
\end{pmatrix}.
\end{equation}

Before turning to a detailed calculation of the electric current let us briefly
address the issue of the spin current conservation. It is worth pointing out that
in general the spin current needs not to be conserved in heterostructures
with spin active interfaces, see, e.g., Ref. \onlinecite{Tserkovnyak05} and
further references therein. Only in certain specific situations such
conservation can take place. For instance, one can easily check that a single
barrier with $S$-matrix \eqref{S22}-\eqref{S22'} preserves the spin current 
conservation \cite{Zhao07}. Making use of the general expressions for
the Green-Keldysh functions we have verified that in our two barrier 
structure with  interface $S$-matrices \eqref{S11}-\eqref{S22'} the 
spin current is in general {\it not} conserved even in the
normal state. For instance, spin accumulation on the first barrier is
controlled by the combination $(D_{2\uparrow} - D_{2\downarrow})
\sin\varphi$ which vanishes only for collinear barrier polarizations or 
spin-isotropic second interface. We see no particular reasons
to expect conservation of the spin current in general NSN structures
with non-collinear interface polarizations. However, this general case 
requires a more detailed analysis which goes beyond the frames of the present
paper and will be published elsewhere.

Now we turn to the analysis of the electric current. 
Substituting the above expressions for the $S$-matrices into Eqs.
\eqref{I11} and \eqref{I12} we arrive at the final results for
both $I_{11}(V_1)$ and $I_{12}(V_2)$ which will be specified
further below.

\subsection{Cross-current}
First let us consider the cross-current $I_{12}(V_2)$. From the above
analysis we obtain
\begin{widetext}
\begin{equation}
\begin{split}
I_{12}(V_2)= -\dfrac{G_0}{4e} \int d \varepsilon &
\left[\tanh\dfrac{\varepsilon+eV_2}{2T} -
\tanh\dfrac{\varepsilon}{2T} \right]
\dfrac{1-\tanh^2iL\Omega/v_F}{W(z_1,z_2,\varepsilon,\varphi)}
\times\\  \times \Bigl\{ & \left[D_{1\downarrow}D_{2\downarrow}-
|a|^2D_{1\uparrow}D_{2\downarrow}(R_{1\downarrow}+R_{2\uparrow})+
|a|^4
D_{1\downarrow}R_{1\uparrow}D_{2\downarrow}R_{2\uparrow}\right]
|K(z_1,z_2,\varepsilon)|^2\cos^2(\varphi/2) +\\ + &
\left[D_{1\uparrow} D_{2\uparrow}
-|a|^2D_{1\downarrow}D_{2\uparrow}(R_{1\uparrow}+R_{2\downarrow})
+|a|^4 D_{1\uparrow}
R_{1\downarrow}D_{2\uparrow}R_{2\downarrow}\right]
|K(z_1^*,z_2^*,\varepsilon)|^2\cos^2(\varphi/2) +\\ + &
\left[D_{1\uparrow}D_{2\downarrow}-
|a|^2D_{1\downarrow}D_{2\downarrow}(R_{1\uparrow}+R_{2\uparrow})+
|a|^4
D_{1\uparrow}R_{1\downarrow}D_{2\downarrow}R_{2\uparrow}\right]
|K(z_1^*,z_2,\varepsilon)|^2\sin^2(\varphi/2) +\\ + &
\left[D_{1\downarrow} D_{2\uparrow}
-|a|^2D_{1\uparrow}D_{2\uparrow}(R_{1\downarrow}+R_{2\downarrow})
+|a|^4 D_{1\downarrow}
R_{1\uparrow}D_{2\uparrow}R_{2\downarrow}\right]
|K(z_1,z_2^*,\varepsilon)|^2\sin^2(\varphi/2) \Bigr\},
\end{split}
\label{I12phi}
\end{equation}
where we define
\begin{gather}
K(z_1,z_2,\varepsilon)= (1-a^2 z_1 z_2) - \left[ \varepsilon (1+a^2 z_1 z_2) + \Delta a (z_1+z_2) \right]Q,
\\
W(z_1,z_2,\varepsilon,\varphi)= |K(z_1,z_2,\varepsilon)
K(z_1^*,z_2^*,\varepsilon) \cos^2(\varphi/2) +
K(z_1^*,z_2,\varepsilon) K(z_1,z_2^*,\varepsilon)
\sin^2(\varphi/2)|^2
\end{gather}
and $z_i=\sqrt{ R_{i\uparrow} R_{i\downarrow} }\exp( i \theta_i)$
( $i=1,2$).
\end{widetext}

Eq.~\eqref{I12phi} represents our central result. It fully
determines the non-local spin-dependent current in our
three-terminal ballistic NSN structure at arbitrary voltages,
temperature, interface transmissions and polarizations.

Let us introduce the non-local differential conductance
\begin{equation}
G_{12}(V_2)=-\dfrac{\partial I_{1}}{\partial V_2}
=-\dfrac{\partial I_{12}(V_2)}{\partial V_2}.
\end{equation}
Before specifying this quantity further it is important to observe
that in general the conductance $G_{12}(V_2)$ is not an even
function of the applied voltage $V_2$. This asymmetry arises due
to formation of Andreev bound states in the vicinity of a
spin-active interface \cite{Fogel00,Barash02}. It disappears
provided the spin mixing angles $\theta_1$ and $\theta_2$ remain
equal to $0$ or $\pi$.

In the normal state we have $I_{12}(V_2)=-G_{N_{12}}V_2$, where
\begin{multline}
G_{N_{12}}= \dfrac{G_0}{2} \bigl[ (D_{1\downarrow}D_{2\downarrow}
+ D_{1\uparrow} D_{2\uparrow})\cos^2(\varphi/2) +\\+
(D_{1\uparrow}D_{2\downarrow} + D_{1\downarrow}
D_{2\uparrow})\sin^2(\varphi/2) \bigr]. \label{NNN}
\end{multline}

Turning to the superconducting state, let us consider the limit of
low temperatures and voltage $T, V_2 \ll \Delta$. In this
limit only subgap quasiparticles contribute to the cross-current and the
differential conductance becomes voltage-independent, i.e.
$I_{12}=-G_{12}V_{2}$, where
\begin{widetext}
\begin{multline}
G_{12}=G_0(1-\tanh^2L\Delta/v_F)
\Biggl\{
\dfrac{D_{1\uparrow}D_{1\downarrow}D_{2\uparrow}D_{2\downarrow}}{
|K(z_1,z_2,0)|^2\cos^2(\varphi/2)+|K(z_1,z_2^*,0)|^2\sin^2(\varphi/2)}
+\\+
(D_{1\uparrow}-D_{1\downarrow})(D_{2\uparrow}-D_{2\downarrow})
\dfrac{|K(z_1,z_2,0)|^2\cos^2(\varphi/2)-|K(z_1,z_2^*,0)|^2\sin^2(\varphi/2)
}{\left(|K(z_1,z_2,0)|^2\cos^2(\varphi/2)+|K(z_1,z_2^*,0)|^2\sin^2(\varphi/2)\right)^2}
\Biggr\}.
\label{G12zeroV}
\end{multline}
\end{widetext}
In the case of spin-isotropic interfaces Eqs. \eqref{G12zeroV} and
\eqref{I12phi} reduce to our previous results \cite{KZ06}.

\begin{figure}
\centerline{
\includegraphics[width=75mm]{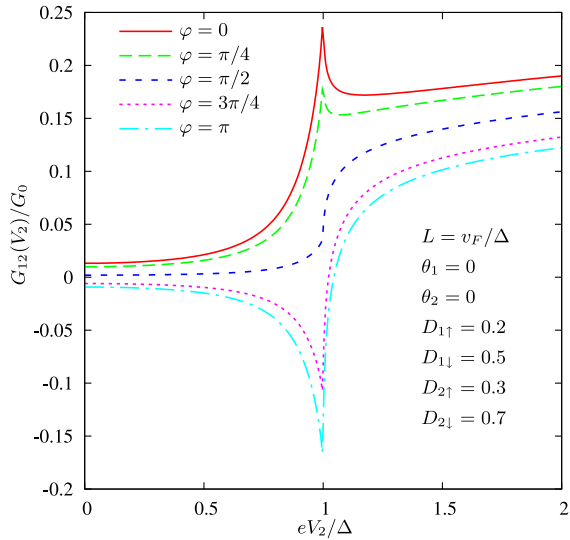}
} \caption{Zero temperature differential non-local conductance as
a function of voltage at zero spin-mixing angles $\theta_{1,2}=0$. }
\label{current-gen-theta=0}
\end{figure}

Provided at least one of the interfaces is spin-isotropic, the
conductance \eqref{G12zeroV} is proportional to the product of all
four transmissions
$D_{1\uparrow}D_{1\downarrow}D_{2\uparrow}D_{2\downarrow}$, i.e.
it differs from zero only due to processes involving scattering
with both spin projections at both NS interfaces. As in the case
of spin-isotropic interfaces \cite{KZ06} the value $G_{12}$
\eqref{G12zeroV} gets strongly suppressed with increasing $L$, and
at sufficiently high interface transmissions this dependence is in
general non-exponential in $L$. In the spin-degenerate case for a
given $L$ the non-local conductance reaches its maximum for
reflectionless barriers $D_{1,2}=1$. In this case we arrive at a
simple formula
\begin{equation}
G_{12}=G_0(1-\tanh^2L\Delta/v_F).
\label{D=1}
\end{equation}
We observe that for small $L \ll v_F/\Delta$ the conductance
$G_{12}$ identically coincides with its normal state value
$G_{N_{12}}\equiv G_0$ at any temperature and voltage \cite{KZ06}.
This result implies that CAR {\it vanishes for fully open
barriers}. Actually this conclusion is general and applies not
only for small but for any value of $L$, i.e. the result
(\ref{D=1}) is determined solely by the process of direct electron
transfer between N-terminals for all $L$.

At the first sight, this result might appear counterintuitive
since the behavior of ordinary (local) AR is just the opposite: It
reaches its maximum at full barrier transmissions. The physics
behind vanishing of CAR for perfectly trasparent NS interfaces is
simple. One observes (cf. Fig. 1) that CAR inevitably implies the
flow of Cooper pairs out of the contact area into the
superconducting terminal. This flow is described by electron
trajectories which end deep in the superconductor. On the other
hand, it is obvious that CAR requires ``mixing'' of these
trajectories with those going straight between two normal
terminals. Provided there exists no normal electron reflection at
both NS interfaces such mixing does not occur, CAR vanishes and
the only remaining contribution to the non-local conductance is
one from direct electron transfer between N-terminals.

This situation is illustrated by the diagrams in Fig. \ref{traject}.
It is obvious that in the case of non-reflecting NS interfaces
only the process of Fig. 5a survives, whereas all other processes
(Fig. 5b, 5c and 5d) vanish for reflectionless barriers with
$R_{1(2)\uparrow(\downarrow)}=0$. The situation changes provided
at least one of the transmissions is smaller than one. In this
case scattering at SN interfaces mixes up trajectories connecting
N$_1$ and N$_2$ terminals with ones going deep into and coming
from the superconductor. As a result, all four processes depicted
in Fig. 5 contribute to the cross-current and CAR contribution to
$G_{12}$ does not vanish.

Let us also note that the statement about the absence of CAR for
highly transparent interfaces was independently made in a recent paper
\cite{DM06}. Although no derivation supporting this statement was
presented, this statement \cite{DM06} appears to be
based on the BTK-like description of strictly 1d NSN structures. 
According to our general analysis, no processes presented in Figs.
5b, 5c and 5d would be possible in that case and, hence, CAR should
be prohibited in 1d NSN systems for {\it any} transmission. This
observation, however, can be easily overlooked if one only deals
with the solutions of the Bogolyubov-de Gennes equations (equivalent
to finding retarded and advanced Green functions) and does
not directly evaluate the electron distribution function (or the Keldysh
function) in the S-terminal.

In the limit $|eV_2|, T \ll \Delta$ and at zero spin-mixing angles
$\theta_{1,2}=0$ from Eq. \eqref{G12zeroV} we obtain
\begin{multline}
G_{12}=G_0\dfrac{1-\tanh^2L\Delta/v_F}{|K(z_1,z_2,0)|^2}
\bigl\{
D_{1\uparrow}D_{1\downarrow}D_{2\uparrow}D_{2\downarrow}
+\\+
(D_{1\uparrow}-D_{1\downarrow})(D_{2\uparrow}-D_{2\downarrow})\cos\varphi
\bigr\}.
\label{G12zeroVzeroTheta}
\end{multline}
In the lowest (first order) order in the transmissions of both
interfaces and for collinear interface polarizations Eq.
\eqref{G12zeroVzeroTheta} reduces to the result by Falci {\it et
al.} \cite{FFH} provided we identify the tunneling density of
states $N_0 D_{1\uparrow}$, $N_0 D_{1\downarrow}$, $N_0
D_{2\uparrow}$, and $N_0 D_{2\downarrow}$ with the corresponding
spin-resolved densities of states in the ferromagnetic electrodes.
For zero spin-mixing angles and low voltages the $L$-dependence of
the nonlocal conductance $G_{12}$ reduces to the exponential form
$G_{12} \propto \exp (-2L\Delta /v_F)$ either in the limit of
small transmissions or large  $L \gg v_F/\Delta$.

\begin{figure}
\centerline{
\includegraphics[width=75mm]{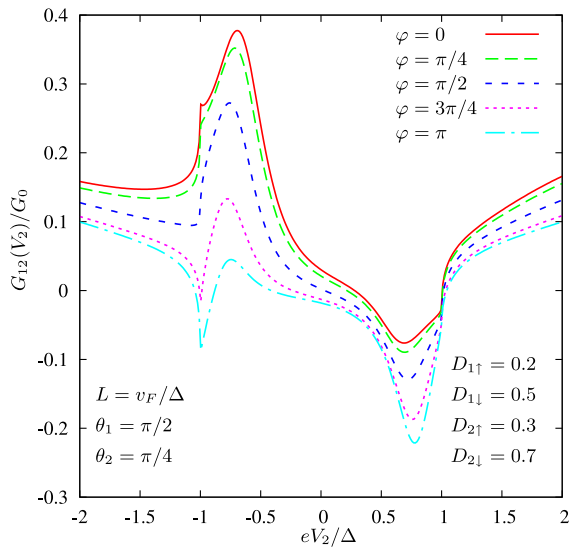}
} \caption{The same as in Fig. 6 for $\theta_1=\pi/2$, $\theta_2=\pi/4$. }
\label{current-gen-theta-1}
\end{figure}

At arbitrary voltages and temperatures the cross-current has a
simple $\varphi$ dependence in the limit of zero spin mixing
angles ($\theta_{1,2}=0$)
\begin{multline}
I_{12}(\varphi,V_2)= I_{12}(\varphi=0,V_2)\cos^2(\varphi/2) +\\+
I_{12}(\varphi=\pi,V_2)\sin^2(\varphi/2), \label{I12theta0}
\end{multline}
i.e. in this limit at any $\varphi$ the nonlocal current is equal
to a proper superposition of the two contributions corresponding
to parallel ($\varphi=0$) and antiparallel ($\varphi=\pi$)
interface polarizations. Some typical curves for the differential
non-local conductance are presented in
Fig.~\ref{current-gen-theta=0} at sufficiently high interface
transmissions and zero spin mixing angles $\theta_{1,2}=0$.

Let us now turn to the limit of highly polarized interfaces which
is accounted for by taking the limit of vanishing spin-up (or
spin-down) transmission of each interface. In this limit our model
describes an HSH structure, where H stands for fully
spin-polarized half-metallic electrodes. In this case we obtain
($D_{1\uparrow}=D_1$, $D_{1\downarrow}=0$, $D_{2\uparrow}=D_2$,
and $D_{2\downarrow}=0$)
\begin{multline}
I_{12}(V_2)=
-\dfrac{G_0}{4e} \int d \varepsilon
\left[h_0(\varepsilon+eV_2) -h_0(\varepsilon)\right]
\times\\\times
\dfrac{1-\tanh^2iL\Omega/v_F}{W(z_1,z_2,\varepsilon,\varphi)}
D_1 D_2
\times\\\times
\Bigl\{
\left[1+|a|^4\right]|K(z_1^*,z_2^*,\varepsilon)|^2\cos^2(\varphi/2)
-\\-
2 |a|^2|K(z_1,z_2^*,\varepsilon)|^2\sin^2(\varphi/2)
\Bigr\}.
\label{hsh}
\end{multline}
We observe that the nonlocal conductance has {\it opposite signs} for
parallel ($\varphi=0$) and antiparallel ($\varphi=\pi$) interface
polarizations. We also emphasize that, as it is also clear from
Eq. (\ref{G12zeroVzeroTheta}), the cross-conductance $G_{12}$
of HSH structures -- in contrast to that for NSN structures --
does not vanish already in the lowest order in barrier transmissions
$D_{1\uparrow}D_{2\uparrow}$.

In general the non-local conductance is very sensitive to
particular values of the spin-mixing angles $\theta_1$ and
$\theta_2$, as illustrated, e.g., in Fig.
\ref{current-gen-theta-1}. Comparing the voltage dependencies of
the nonlocal conductance evaluated for the same transmissions and
presented in Figs. \ref{current-gen-theta=0} and
\ref{current-gen-theta-1}, we observe that they can differ
drastically at zero and non-zero values of $\theta_{1,2}$.

At low voltages and temperatures and at zero spin mixing angles
the non-local conductance of HSH structures is determined by Eq.
(\ref{G12zeroVzeroTheta}) with
$D_{1\downarrow}=D_{2\downarrow}=0$. For fully open barriers (for
''spin-up'' electrons) $D_{1\uparrow}=D_{2\uparrow}=1$ we obtain
\begin{equation}
G_{12}=G_0(1-\tanh^2L\Delta/v_F)\cos \varphi . \label{DD=1}
\end{equation}
Interestingly, for $\varphi =0$ this expression exactly coincides
with that for fully open NSN structures, Eq. (\ref{D=1}). At the
same time for small $L$ the result (\ref{DD=1}) turns out to be 2
times bigger that the analogous expression in the normal case, i.e. for
fully open HNH structures, cf. Eq. (\ref{NNN}). This result can 
easily be interpreted in terms of diagrams in Fig. 5. We observe that
-- exactly as for the spin degenerate case -- CAR diagrams of Fig.
5b,c vanish for reflectionless barriers, whereas diagrams of
Fig. 5a,d describing direct electron transfer survive and both
contribute to $G_{12}$. Thus, {\it CAR vanishes identically also
for fully open HSH structures}. The factor of 2 difference with
the normal case is due to the fact that the diagram of Fig. 5d
vanishes in the normal limit.

Finally, let us  compare our general results, Eq. \eqref{I12phi}
and below, with the the analogous results \cite{MF}. In order to
account for higher order tunneling events in the case of symmetric
interfaces the authors \cite{MF} employed effective ''dressing''
of both EC and CAR contributions by all higher order {\it local}
Andreev processes at both interfaces. At the same time, all higher
order {\it non-local} processes were ignored in Ref.
\onlinecite{MF}. Our analysis -- which includes all processes to
all orders -- demonstrates that this approximation \cite{MF} may
be appropriate only in the limit $L \gg v_F/ \Delta$ in which case
non-local effects are exponentially suppressed. Otherwise higher
order non-local processes remain important and need to be fully
accounted for.

Comparing our exact results and those of Ref. \onlinecite{MF}
(with omitted higher order non-local processes) we observe that
the disagreement between them grows with increasing transmissions
and becomes maximal at full transmissions and small values of $L$.
The most significant differences are: (i) the dependence of
$G_{12}$ on $L$ is non-exponential whereas in Ref. \onlinecite{MF}
it remains exponential at all $L$, (ii) at small $L$ and high
transmissions the linear conductance $G_{12}$ evaluated in Ref.
\onlinecite{MF} (see e.g. Eqs. (33), (34) of that work) turns out to be
several times smaller than our results in both cases of NSN and
HSH structures and (iii) taking into account only local Andreev
reflection events is not sufficient to correctly account for
vanishing CAR in the limit of fully open interfaces.

\subsection{Correction to BTK}
Using the above formalism one can easily generalize the BTK result
to the case of spin-polarized interfaces\cite{Zhao04}.
For the first interface we have
\begin{multline}
I_1^{BTK}(V_1)= \dfrac{\mathcal{N}_1}{R_qe} \int d \varepsilon
[h_0(\varepsilon+eV_1)-h_0(\varepsilon)] (1+|a|^2) \times\\\times
\left< \dfrac{|v_{x_1}|}{v_F} \left(
D_{1\uparrow}\dfrac{1-R_{1\downarrow}|a|^2}{|1-z_1 a^2|^2} +
D_{1\downarrow}\dfrac{1-R_{1\uparrow}|a|^2}{|1-z_1^* a^2|^2}
\right) \right>.
\end{multline}
Here transmission and reflection coefficients as well as the spin
mixing angle depend on the direction of the Fermi momentum. In the
spin-degenerate case the above expression reduces to the standard
BTK result \cite{BTK}.

Evaluating the nonlocal correction to the BTK current due to the
presence of the second interface we arrive at a somewhat lengthy
general expression
\begin{widetext}
\begin{gather}
\begin{split}
I_{11}(V_1)=
\dfrac{G_0}{2e} \int d \varepsilon
(h_0(\varepsilon+eV_1)-h_0(\varepsilon))
\dfrac{1}{W(z_1,z_2,\varepsilon,\varphi)}\Bigl\{ & 2W(z_1,z_2,\varepsilon,\varphi)
-\\-
R_{1\uparrow}
\bigl| \cos^2(\varphi/2) K(z_1/R_{1\uparrow}, z_2,\varepsilon) K(z_1^*, z_2^*,\varepsilon) + &
\sin^2(\varphi/2)K(z_1/R_{1\uparrow}, z_2^*,\varepsilon) K(z_1^*, z_2,\varepsilon) \bigr|^2
-\\-
R_{1\downarrow}
\bigl| \cos^2(\varphi/2) K(z_1^*/R_{1\downarrow}, z_2^*,\varepsilon) K(z_1, z_2,\varepsilon) + &
\sin^2(\varphi/2)K(z_1^*/R_{1\downarrow}, z_2,\varepsilon) K(z_1, z_2^*,\varepsilon) \bigr|^2
\Bigr\}+
\end{split}
\notag
\\
\begin{split}
+\dfrac{G_0}{4e} \int d \varepsilon
(h_0(\varepsilon+eV_1)-h_0(\varepsilon))
\dfrac{D_{1\uparrow}D_{1\downarrow}}{W(z_1,z_2,\varepsilon,\varphi)}\Bigl\{&
\\
|a|^2 \bigl| \cos^2(\varphi/2) K(0,z_2,\varepsilon) K(z_1^*, z_2^*,\varepsilon) + &
\sin^2(\varphi/2)K(0,z_2^*,\varepsilon) K(z_1^*, z_2,\varepsilon) \bigr|^2
+\\+
|a|^2 \bigl| \cos^2(\varphi/2) K(0,z_2^*,\varepsilon) K(z_1, z_2,\varepsilon) + &
\sin^2(\varphi/2)K(0,z_2,\varepsilon) K(z_1, z_2^*,\varepsilon) \bigr|^2
+\\+
\dfrac{1}{|a|^2}
\bigl| \cos^2(\varphi/2) K'(z_2^*,\varepsilon) K(z_1, z_2,\varepsilon) + &
\sin^2(\varphi/2)K'(z_2,\varepsilon) K(z_1, z_2^*,\varepsilon) \bigr|^2
+\\+
\dfrac{1}{|a|^2}
\bigl| \cos^2(\varphi/2) K'(z_2,\varepsilon) K(z_1^*, z_2^*,\varepsilon) + &
\sin^2(\varphi/2)K'(z_2^*,\varepsilon) K(z_1^*, z_2,\varepsilon) \bigr|^2
\Bigr\}+
\end{split}
\notag
\\+
\dfrac{G_0}{e} R_{2\uparrow}R_{2\downarrow}\sin^2(\theta_2/2)
\sin^2(\varphi/2)\cos^2(\varphi/2)
\int d \varepsilon
(h_0(\varepsilon+eV_1)-h_0(\varepsilon))
\dfrac{(1-\tanh^2iL\Omega/v_F)^2
}{W(z_1,z_2,\varepsilon,\varphi)}
\times
\notag
\\\times
\left[|a|^2(D^2_{1\uparrow}+D^2_{1\downarrow}) - 2 |a|^4
D_{1\uparrow}D_{1\downarrow}(R_{1\uparrow} + R_{1\downarrow}) +
|a|^6
(D^2_{1\uparrow}R^2_{1\downarrow}+D^2_{1\downarrow}R^2_{1\uparrow})\right],
\label{CARcorr}
\end{gather}
\end{widetext}
where $K'(z_2,\varepsilon)=\partial K(z_1,
z_2,\varepsilon)/\partial z_1$. This expression gets significantly
simplified in the limit of zero spin-mixing angles $\theta_{1,2}=
0$ in which case we obtain
\begin{multline}
I_{11}(V_1)=
\dfrac{G_0}{2e} \int d \varepsilon
(h_0(\varepsilon+eV_1)-h_0(\varepsilon))
\\
\Biggl\{
2-
R_{1\uparrow}
\dfrac{|K(z_1/R_{1\uparrow}, z_2,\varepsilon)|^2}{|K(z_1,z_2,\varepsilon)|^2}
-
R_{1\downarrow}
\dfrac{|K(z_1/R_{1\downarrow}, z_2,\varepsilon)|^2}{|K(z_1,z_2,\varepsilon)|^2}
+\\+
D_{1\uparrow}D_{1\downarrow}
\dfrac{
|a(\varepsilon)|^2 |K(0,z_2,\varepsilon)|^2+
|K'(z_2,\varepsilon)|^2/|a(\varepsilon)|^2}{|K(z_1,z_2,\varepsilon)|^2}
\Biggr\}.
\end{multline}
In contrast to the expression for the cross-current $I_{12}$ (cf.
Eq. \eqref{I12theta0}), in the limit of zero spin-mixing angles
the correction $I_{11}$ to the BTK current does not depend on the
angle $\varphi$ between the interface polarizations. In
particular, at $|eV_1|, T \ll \Delta$ we have $I_{11}= G_{11}V_1$
where
\begin{multline}
G_{11}=
G_0
(D_{1\uparrow}+D_{1\downarrow})
\dfrac{(1-z_2^2)(1-\tanh^2L\Delta/v_F)}{
[1+z_1 z_2 +(z_1+z_2 )\tanh L\Delta/v_F]^2}
+\\+
G_0
D_{1\uparrow}D_{1\downarrow}
\dfrac{(1+z_2 \tanh L\Delta/v_F)^2 + 3 (z_2 + \tanh L\Delta/v_F)^2}{[1+z_1 z_2 +(z_1+z_2 )\tanh L\Delta/v_F]^2}.
\end{multline}
In the tunneling limit
$D_{1\uparrow},D_{1\downarrow},D_{2\uparrow},D_{2\downarrow} \ll
1$ we reproduce the result of Ref. \onlinecite{FFH}
\begin{equation}
G_{11}=\dfrac{G_0}{4}
(D_{1\uparrow}+D_{1\downarrow})(D_{2\uparrow}+D_{2\downarrow})
\exp(-2L\Delta/v_F),
\end{equation}
which turns out to hold at any value $\varphi$.

As compared to the BTK conductance the CAR correction
\eqref{CARcorr} contains an extra small factor ${\cal A}_2/L^2$
and, hence, in many cases remains small and can be neglected. On
the other hand, since CAR involves tunneling of {\it one} electron
through each interface, for strongly asymmetric structures with
$D_{1\uparrow},D_{1\downarrow} \ll 1$ and $D_{2\uparrow},
D_{2\downarrow} \sim 1$ it can actually {\it strongly exceed} the
BTK conductance. Indeed, for $D_{1\uparrow\downarrow} \ll 1$,
$R_{2\uparrow}R_{2\downarrow} \ll 1$ and provided the spin mixing
angle $\theta_1$ is not very close to $\pi$ from Eq.
\eqref{CARcorr} we get
\begin{equation}
G_{11}= \dfrac{G_0 (D_{1\uparrow}+D_{1\downarrow})}{\cosh (2
L\Delta/v_F) + \cos \theta_1 \sinh (2 L\Delta/v_F)},
\label{CARcorr1}
\end{equation}
i.e. for
$$
\frac{D_{1\uparrow} D_{1\downarrow}}{(D_{1\uparrow} +
D_{1\downarrow})} < \frac{{\cal A}_2}{L^2}\exp(-2L\Delta/v_F)
$$
the contribution \eqref{CARcorr1} may well exceed the BTK term
$G_{1}^{BTK} \propto D_{1\uparrow} D_{1\downarrow}$. The existence
of such a non-trivial regime further emphasizes the importance of
the mechanism of non-local Andreev reflection in multi-terminal
hybrid NSN structures.

\section{Conclusions}

In this paper we developed a non-perturbative theory of non-local
electron transport in ballistic NSN three-terminal structures with
spin-active interfaces. Our theory is based on the quasiclassical
formalism of energy-integrated Green-Eilenberger functions
supplemented by appropriate boundary conditions describing
spin-dependent scattering at NS interfaces. Our approach applies
at arbitrary interface transmissions and allows to fully describe
non-trivial interplay between spin-sensitive normal scattering,
local and non-local Andreev reflection at SN interfaces. Our main
results are the general expressions for the non-local
cross-current $I_{12}$, Eq. \eqref{I12phi}, and for the non-local
correction $I_{11}$ to the BTK current, Eq. \eqref{CARcorr}. These
expressions provide complete description of the conductance matrix
of our three-terminal NSN device at arbitrary voltages,
temperature, spin-dependent transmissions of NS interfaces and
their polarizations.

Our analysis allows to predict and analyze a rich variety of
interesting properties of such structures. One of our predictions
is that in the case of ballistic electrodes no crossed Andreev
reflection can occur in both NSN and HSH structures with fully
open interfaces. Beyond the tunneling limit the dependence of the
non-local conductance on the size of the S-electrode $L$ is in
general non-exponential and reduces to $G_{12} \propto \exp
(-2L\Delta /v_F )$ only in the limit of large $L$. For hybrid
structures half-metal-superconductor-half-metal we predict that
the low energy non-local conductance does not vanish already in
the lowest order in barrier transmissions $G_{12} \propto
D_{1\uparrow}D_{2\uparrow}$. These and other our predictions can
be directly tested in future experiments on NSN hybrid structures,
including systems with ferromagnetic and half-metallic electrodes.

\section*{Acknowledgments}

We are indebted to D. Beckmann for pointing out numerical mistakes in initial
versions of Fig. 6 and 7.
This work is part of the EU Framework Programme
NMP4-CT-2003-505457 ULTRA-1D "Experimental and theoretical investigation of
electron transport in ultra-narrow 1-dimensional nanostructures".


\begin{thebibliography}{99}
\bibitem{FN} The presence of an additional dot
with vanishing superconducting order parameter \cite{Belzig} can
cause modifications of the physical picture of non-local electron
transport as compared to that analyzed in other works.
\end{thebibliography}
\end{document}